\documentclass[a4paper,preprint,showkeys,superscriptaddress]{revtex4}
\usepackage{epsfig}

\begin{document}
\title{Scintillator detectors with long WLS fibers and multi-pixel photodiodes}

\author{O.~Mineev}\email[Corresponding author. E--mail: ]{oleg@inr.ru}\affiliation{Institute for Nuclear Research of RAS}
\author{Yu.~Kudenko}\affiliation{Institute for Nuclear Research of RAS}  
\author{Yu.~Musienko}\affiliation{Institute for Nuclear Research of RAS}, 
\author{I.~Polyansky}\affiliation{Institute for Nuclear Research of RAS}\affiliation{Center of Perspective Technology and Apparatus,107076 Moscow, Russia}
\author{N.~Yershov}\affiliation{Institute for Nuclear Research of RAS}

\begin{abstract} 
We have studied the possibility of using Geiger mode
multi-pixel photodiodes to read out long scintillator bars
with a single wavelength-shifting fiber embedded  along the bar. This detector configuration can be
used in large volume detectors in future long baseline neutrino
oscillation experiments. Prototype bars of 0.7~cm thickness and different
widths have been produced  and tested using two types of multi-pixel
photodiodes: MRS APD (CPTA, Moscow) and MPPC (Hamamatsu). A minimum light yield of 7.2 p.e./MeV was obtained for a 4~cm wide bar.
\end{abstract}

\maketitle
\section{Introduction}
Scintillator bars with wavelength-shifting (WLS) fibers and opto-electronic readout are considered  an established technology  for massive neutrino tracking calorimeters in  long-baseline neutrino oscillation experiments. The MINOS experiment~\cite{minos} employs extruded bars of $1\times4.1\times800~{\rm cm}^3$ size with 9~m long WLS fibers.  The SciBar detector~\cite{scibar} in the K2K experiment was built with the same technology as MINOS but using shorter bars of $1.3\times2.5\times300~{\rm cm}^3$ size and 3.6~m long WLS fibers of 1.5~mm diameter.  A fine-grained detector in the Miner$\nu$a experiment~\cite{minerva} is made of triangular-shaped 3.5~m long strips and WLS fibers of 1.2~mm diameter. All of these detectors use multi-anode PMTs for optical readout. 

The neutrino far detector of the ``off-axis superbeam'' experiment No$\nu$a~\cite{nova} will be composed of liquid scintillator encased in 15.5~m long  rigid PVC extrusion cells. The scintillator cell is readout by a 30~m long U-shaped WLS fiber of 0.7~mm diameter into avalanche photodiodes (APDs), similar to the APDs developed for the CMS detector.  These devices  have higher quantum efficiency than photomultipliers, but low gain and high intrinsic  noise.
  
A magnetized  iron and scintillator sampling calorimeter  with  a fiducial mass of about 100~kt  and a cross sectional  area of $\sim 14\times14$~m$^2$  with a 1 T dipole field is considered as a baseline option for a future Neutrino Factory detector. An alternative option  is a magnetized, totally active highly segmented scintillator detector of about the same cross sectional area~\cite{iss_det}. Both detectors will consist of a large number of readout channels that require the usage of very compact, insensitive to magnetic field photosensors with a high efficiency to the green light emitted from  WLS fibers. Multi-pixel Geiger mode avalanche photodiodes  are considered as a possible optical readout in these detectors. Detailed information about such devices and their basic principle of operation can be found in ref.~\cite{renker}.  The first application of such photosensors in a large experiment has been done in the near neutrino detector~\cite{nd280} of the long baseline experiment T2K~\cite{t2k} where approximately 56000 Multipixel Photon Counters (MPPCs)~\cite{mppc} are used.

The real challenge lies in the required fine granularity and size of the detectors in these new experiments. Each individual element should provide the ability to detect  minimum ionizing particles with high efficiency in such a large detector system.  In this paper, results of measurements using scintillator detectors  read out with long  WLS fibers  and  Geiger mode multi-pixel avalanche  photodiodes  are presented.

\section{Photosensor spectral sensitivity}
Good spectral matching of the light signal and a photosensor quantum efficiency  is one of the most important factors needed to obtain a high light yield in scintillator detectors. We measured the spectral transmittance of Y11 Kuraray WLS fibers as well as the spectral sensitivities of two types of multi-pixel Geiger photodiodes: Hamamatsu MPPC~\cite{mppc} and MRS APD~\cite{mrs}.

\begin{figure}[htb]
\centering\includegraphics[width=0.8\textwidth]{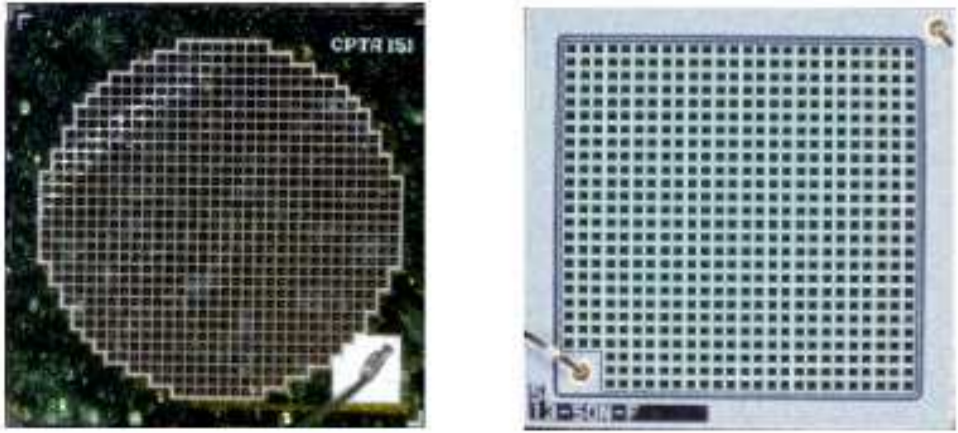}
\caption{Front views of MRS APD (left) and MPPC (right).}
\label{fig:frontview}
\end{figure}

A 796-pixel MRS APD (type CPTA 151-30) is produced by the CPTA company (Moscow, Russia). The sensitive area of this device is approximated by a circle of diameter 1.28~mm. The pixel size is  $43\times 43~\mu$m$^2$. A typical gain is close to $10^6$, the combined crosstalk and afterpulse probability is estimated to be about 10\%,  and the dark rate is  1.2-1.5~MHz.  A customized 667-pixel MPPC (type S10362-13-050C) with an active area of $1.3\times 1.3$ mm$^2$ was developed by Hamamatsu for the near neutrino detector of the T2K experiment. The pixel  size of the MPPC is  $50\times 50~\mu$m$^2$. It should be noted that the response of multi-pixel Geiger photodiodes   depends on  the overvoltage 
$\Delta V$, i.e. excess of bias voltage over the avalanche breakdown value. MPPCs biased at a typical overvoltage of $\Delta V = 1.2$~V at T=25$^\circ$C are characterized by the following parameters:  a gain of $7 \times1 0^5$; an average dark rate of 700~kHz; measured crosstalk and afterpulse probabilities of 9-12\% and 14-16\% respectively, with a combined value of 20-25\%.  A front view of both photosensors is shown in Fig.~\ref{fig:frontview}.

A spectrophotometer calibrated with a PIN-diode was used to measure the photon detection efficiency (PDE) of both MRS APD and MPPC~\cite{musienko}. The PDE spectra were measured at  $\Delta V$=1.2~V for the MPPC and  $\Delta V$=2~V for the MRS APD.  The spectrophotometer light intensity was reduced until the maximum photosensor current was only $\sim$30\%  greater than the dark current of a photodiode to avoid nonlinearity effects caused by the limited number of pixels. Comparing the  measured current  with the calibrated PIN-diode photocurrent we obtain the relative spectral sensitivity. Then the spectral response was normalized using  the reference PDE points obtained with LED's at 410 and 515~nm. Average number of photons in LED pulses was measured using a calibrated PMT XP2020.
The spectral sensitivity could then be expressed in absolute values without the crosstalk, afterpulse and dark rate contributions. 
\begin{figure}[htb]
\centering\includegraphics[width=0.8\textwidth]{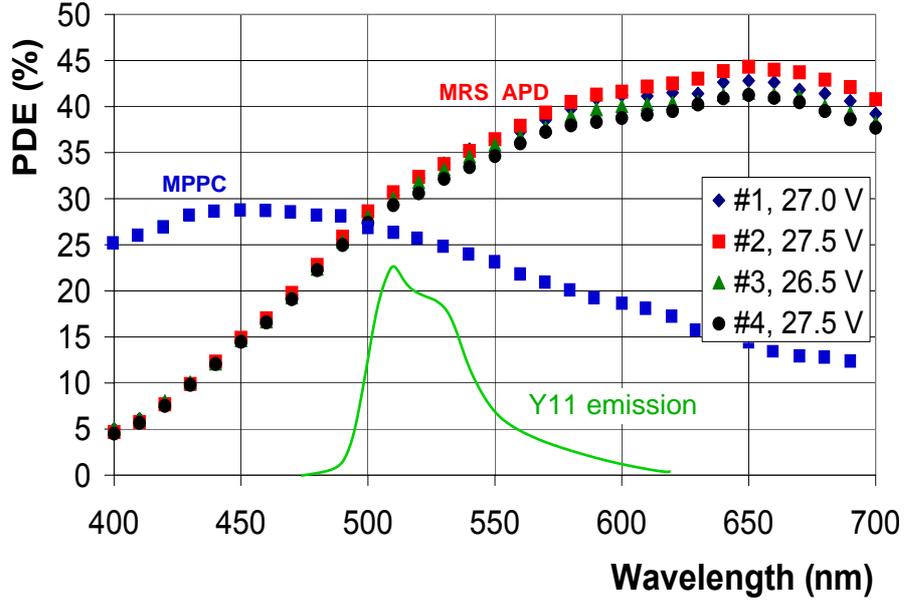}
\caption{Photon detection efficiency  as a function of wavelength for 4 samples of MRS APDs and an MPPC at 25$^\circ$C. The bias voltage for each MRS APD was set to an overvoltage of $\Delta V$=2~V for all tested samples. MPPC was measured at $\Delta V$=1.2~V. Also shown is the Y11(150) Kuraray fiber emission spectrum (in arbitrary units) for a fiber length of 150 cm (Kuraray specification).}
\label{fig:mrsspec}
\end{figure}
The results are shown in  Fig.~\ref{fig:mrsspec}. Both MRS APD and the MPPC have similar values of PDE at the Y11 emission peak of 515~nm measured using a  150~cm long Y11 fiber. However,  the maximum sensitivity of the MRS APDs is shifted to the red wavelengths around 650~nm while the MPPC  sensitivity peaks in the blue region around 450~nm.  

The different spectral sensitivities  of the MRS APD and MPPC within the visible light range can be explained by their different semiconductor layer structures. The MRS APD uses a $n^+$-$p$-$p^+$ structure while  the simplified scheme from the Hamamatsu MPPC catalogue shows the $p^+$-$p^-$-$n^{++}$ structure which is usually associated with a peak spectral sensitivity for blue photons.

\section{Long fiber study }

\subsection{Y11 fiber transmittance}
The transmittance of 1 mm diameter multi-clad Kuraray WLS Y11(200) S-type fibers~\cite{kuraray} was measured for wavelengths of 340-800~nm     using   clear fibers of 1~mm diameter as a reference. The fiber ends were glued in ferrules and polished. Monochromatic light at a selected wavelength was injected into the test fiber through one end. The intensity of the transmitted light  was then measured  by a calibrated photodiode attached to  the opposite end. 

\begin{figure}[htbp]
\centering\includegraphics[width=16cm,angle=0]{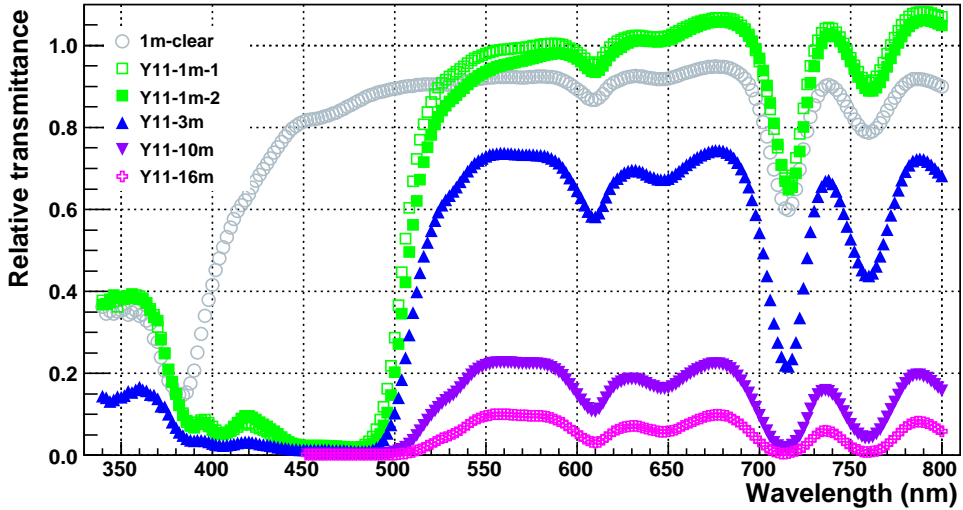}
\caption{Relative spectral transmittance    of 1~m 
(two samples), 3~m, 10~m and 15~m long Y11 Kuraray fibers.   Transmittance is  equal to 1.0 at all wavelengths for a  10~cm long reference clear fiber.}
\label{fig:trans1} 
\end{figure}

\begin{figure}[htbp]
\centering\includegraphics[width=12cm,angle=0]{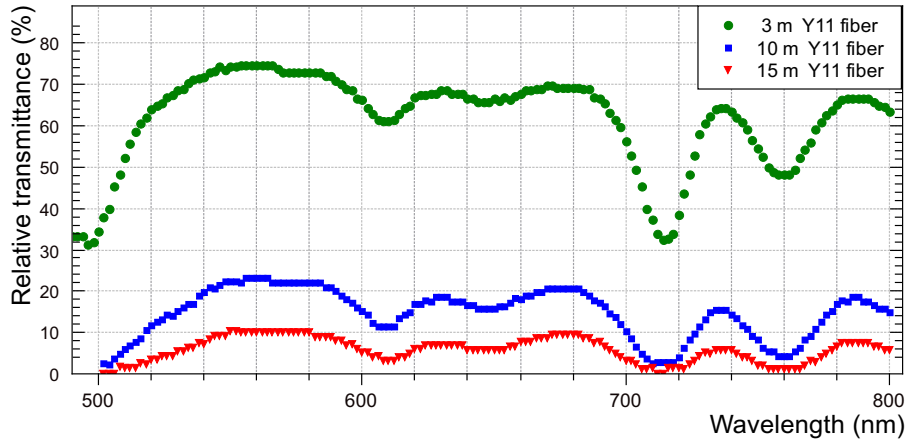}
\caption{Relative spectral transmittance for WLS Y11 Kuraray fibers of different lengths. Transmittance is  equal to  100\% for a 1~m long reference Y11 fiber.}
\label{fig:trans2} 
\end{figure}

The transmittances of Y11 fibers normalized to the transmittance of a 10 cm reference clear fiber are shown in Fig.~\ref{fig:trans1}.
Better transmittance of 1~m long Y11  fibers  than that of the clear fiber of the same length could be explained by some defects or impurities in the tested clear fibers. Fig.~\ref{fig:trans2} 
shows the transmittance of long Y11 fibers relative to a  1~m long Y11 fiber in the 500-800~nm range. A 15~m long WLS fiber absorbs essentially all the light with wavelengths below 500~nm.  For longer wavelengths, the transmittance increases to a maximum  level of about 10\% at 550~nm. Since the Y11 emission spectrum has wavelengths of  500-550~nm  with a low intensity tail extended to over 600~nm (see Fig.~\ref{fig:mrsspec}), one can expect that attenuation of the Y11 signal will be about 95\% after the light travels a distance of 15~m.  

\begin{figure}[htbp]
\centering\includegraphics[width=11cm,angle=0]{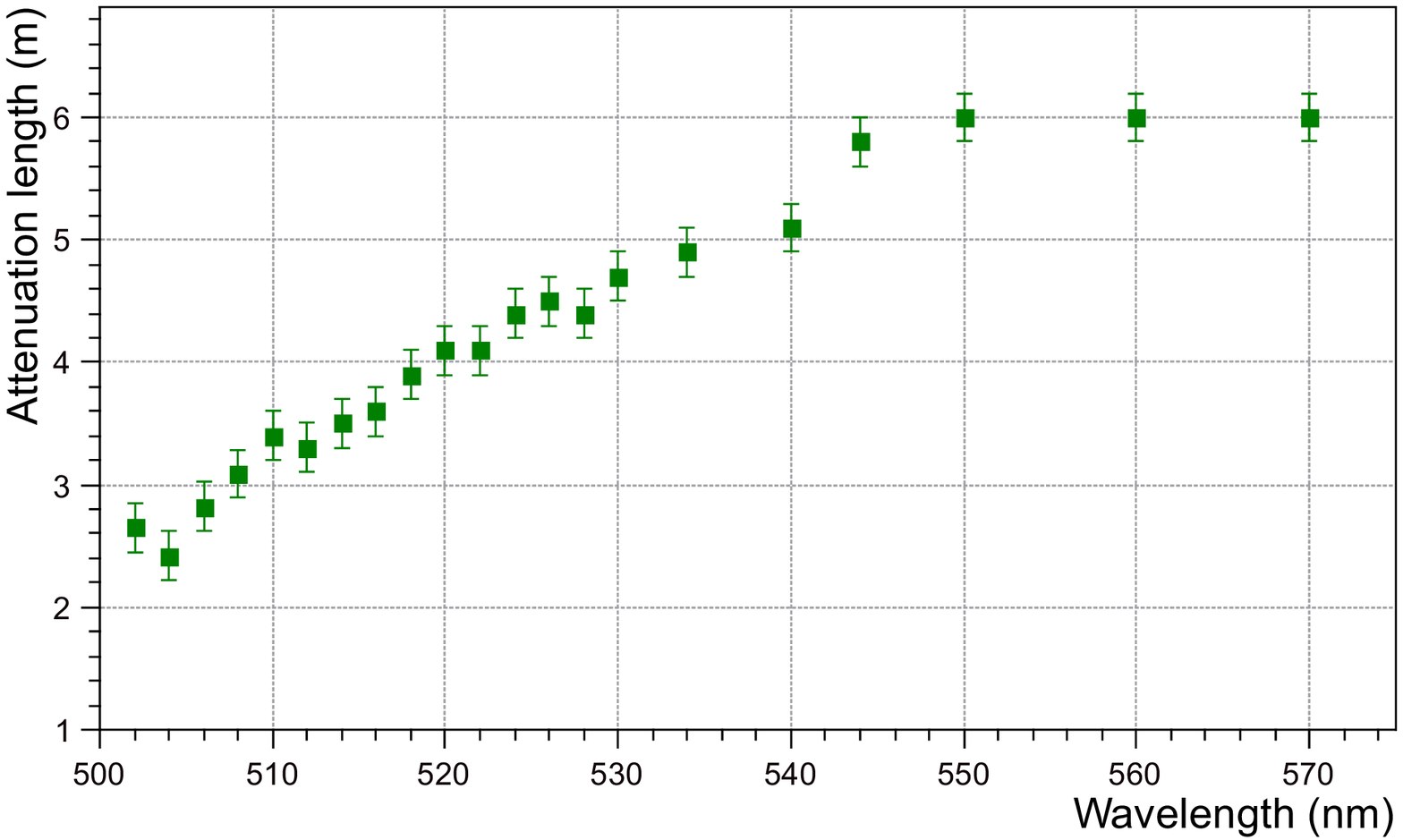}
\caption{The  attenuation length of Y11 fiber vs wavelength.}
\label{fig:att_vs_wv} 
\end{figure}

The dependence of the attenuation length of the Y11 fiber on the wavelength was obtained 
from the transmittance measurements using a single exponential fit function. Fig.~\ref{fig:att_vs_wv} 
 shows the Y11 attenuation length for wavelengths larger than 500~nm. For shorter wavelengths,  the  light signal was too small in 10~m and 15~m long fibers  to obtain reliable values.  

\subsection{Y11 fiber attenuation }
\begin{figure}[htb]
\centering\includegraphics[width=16cm,angle=0]{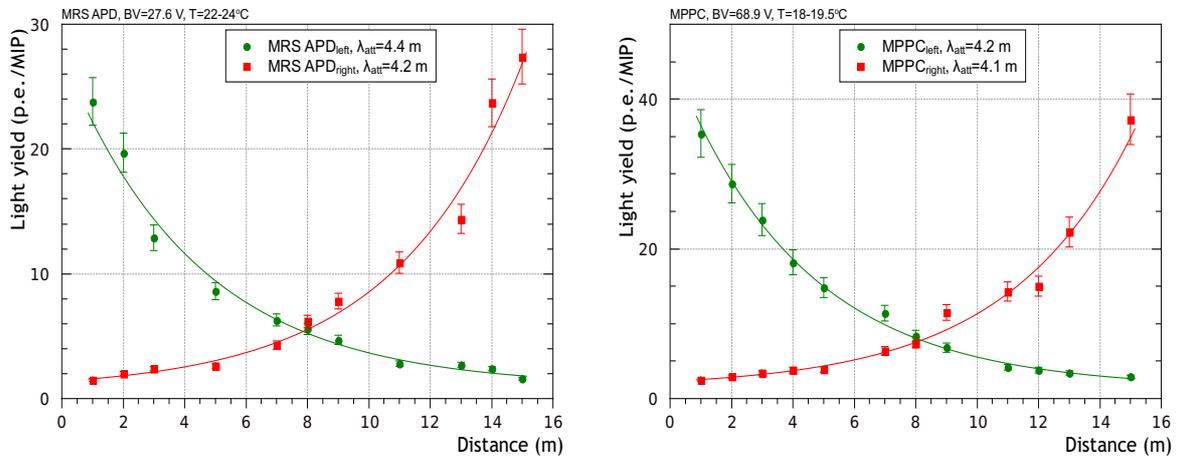}
\caption{Light yield for a minimum ionizing particle (MIP) along the 16~m long Y11 fiber where both fiber ends are read out with either MRS APDs (left plot) or MPPCs (right plot). Attenuation is extracted using a single exponential functional fit.}
\label{fig:Att} 
\end{figure}

To study the attenuation of the re-emitted light  in a 16~m long Y11 fiber,  we carried out measurements of the light yield from cosmic ray muons using a small plastic counter.   The fiber was embedded with optical grease into a groove machined on the scintillator surface and both ends of the fiber were readout by multi-pixel photodiodes.  Photodiode signals  were split  after the preamplifiers and sent to a charge-integrating LeCroy ADC 2249 and discriminators. The ADC gate was set to 200~ns to ensure that the signal was well inside the integration gate after travelling the 16~m fiber length.  The scintillator was moved along the fiber using 1~m steps.   The light yield as a function of distance from the photosensors   is shown in Fig.~\ref{fig:Att}.

A single exponential function was used to fit for the light yield attenuation along the fiber.
An attenuation length of 4.4~m was obtained for the left MRS APD and 4.2~m  for the right one. Attenuation lengths of 4.1 and 4.2~m were obtained for the fiber readout with MPPCs.  The shorter attenuation length with MPPCs was expected because  the MPPC spectral sensitivity peaks in the blue region. However the difference is rather small and within the accuracy ($\sigma = 0.4$ m) of the fit.
An average attenuation length of 4.2~m corresponds to a light attenuation at an effective wavelength of 520~nm  as can be seen in Fig.~\ref{fig:att_vs_wv}.

\begin{figure}[htb]
\centering\includegraphics[width=10cm,angle=0]{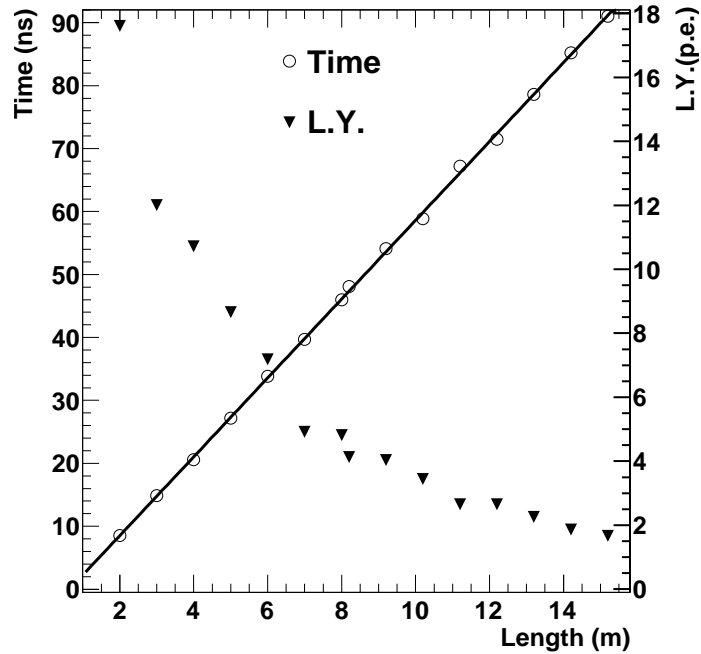}
\caption{ Time difference between the trigger signal and the signal from the MRS APD along a 16~m long Y11 fiber of 1~mm diameter. The timing error of $\pm$0.3~ns is within the radius of the open circles. The light yield in photoelectrons per a minimum ionizing particle is also plotted.}
\label{fig:light_fiber} \end{figure}

To measure the velocity of re-emitted light propagation  in a Y11 fiber, a start signal was generated by a  scintillator  counter from a cosmic ray trigger setup. Another small scintillator tile excited  the fiber, and an MRS APD at one fiber end produced the stop signal. The discriminator threshold was set to be fired by 1~photoelectron (p.e.) pulses. The resolution of the LeCroy TDC 2228A  was calibrated to be $50\pm0.25$~ps/ch.  The time difference between the trigger signal and the signal from the MRS APD  is shown in Fig.~\ref{fig:light_fiber}. The velocity of the re-emitted light propagation in the Y11 fiber is found to be 16.00$\pm 0.08$~cm/ns. 

\section{Measurements of scintillator bars}
Scintillator slabs of $0.7\times20\times90$ cm$^3$ size  were extruded  at the Uniplast Factory (Vladimir, Russia) and then cut to  90~cm long bars with widths of 1, 2, 3 and 4~cm. The scintillator composition is  polystyrene doped with 1.5\% of paraterphenyl (PTP) and 0.01\% of POPOP. The bars were covered by a chemical reflector by etching the scintillator surface  in a chemical agent that results in the formation of a white micropore deposit over the polystyrene~\cite{extrusion} surface.  The chemical coating is an excellent reflector and in addition it smooths out the rough surface acquired during the cutting process.
A 2~mm deep, 1.1~mm wide groove was machined along the center of the bar to accomodate a 16~m long Y11 fiber. Since the tested bar is moved along the fiber, optical grease was used as the coupling between the fiber and the bar.

\begin{figure}[htb]
\centering\includegraphics[width=12cm,angle=0]{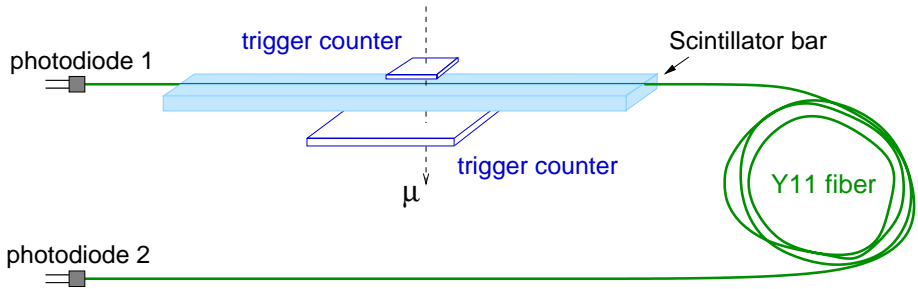}
\caption{Test bench for study of the detector read out using a 16~m long Y11 fiber.}
\label{fig:setup} 
\end{figure}

The test bench for detector measuruments is shown in Fig~\ref{fig:setup}.
A 16~m long fiber was coiled inside a light tight box with a minimum bending radius of about 30~cm. The ionization area within the bars was localized to a $2\times2$ cm$^2$ spot defined by the trigger counter size.  The position of the  trigger counter was fixed to the center of the 90~cm long bar to optimize scintillation light collection by the fiber.  Since a measurement of the 1~cm thick extruded slab yielded an effective attenuation length of the scintillation light in the scintillator of approximately 8.1~cm~\cite{smrd}, we estimate that the full scintillation light collection by a WLS fiber occurs within $\pm$25~cm from the cosmic ray  ionization point.  

\begin{figure}[htb]
\centering\includegraphics[width=12cm,angle=0]{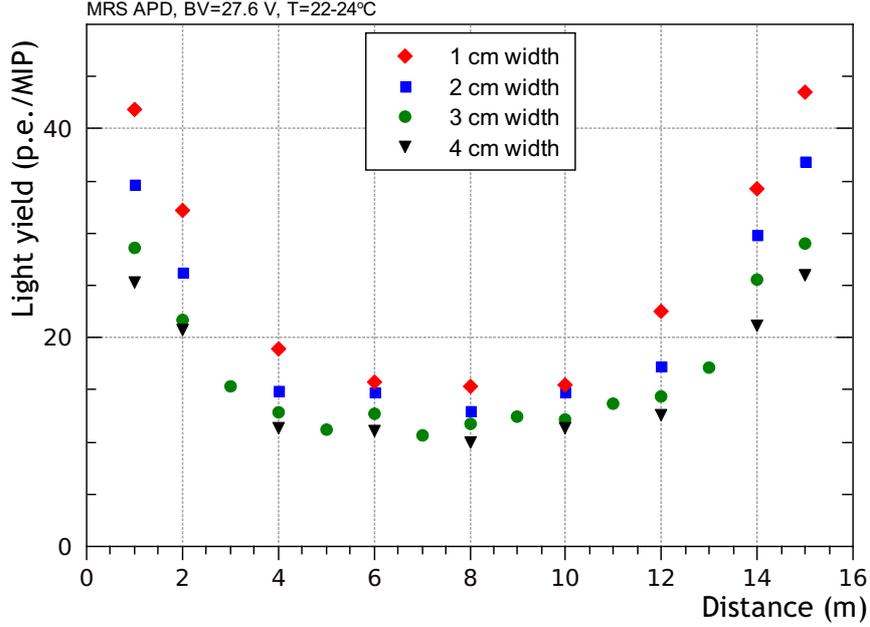}
\caption{Total light yield from both fiber ends  vs the position along the Y11 fiber for 0.7~cm thick bars of different widths. Measurement errors are estimated to be about 7\%.}
\label{fig:bar4} 
\end{figure}

\begin{figure}[htb]
\centering\includegraphics[width=12cm,angle=0]{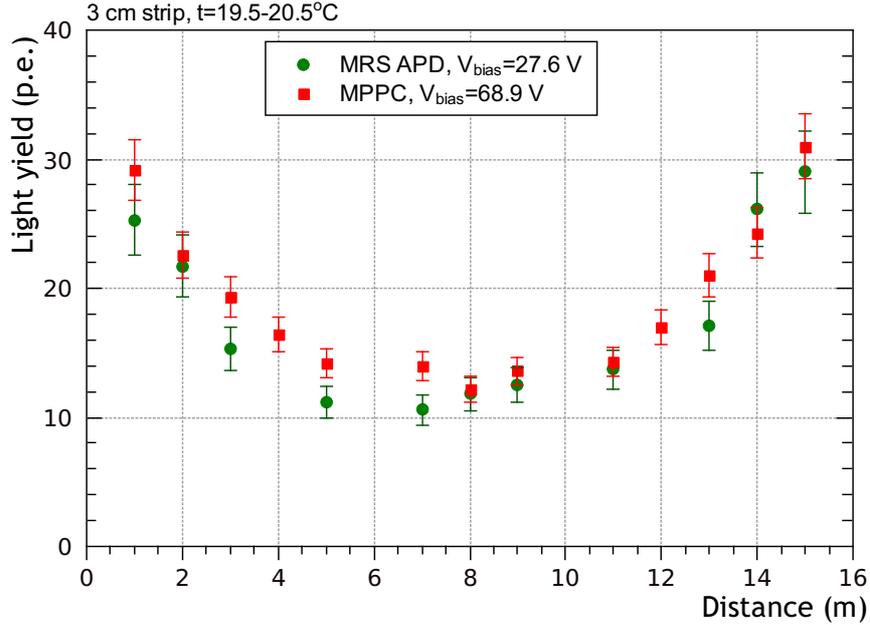}
\caption{Light yield (sum of both  ends) along the Y11 fiber for the 3~cm wide bar read out with MRS APDs and MPPCs.}
\label{fig:bar2} \end{figure}

The  light yield along the fiber  (sum of both ends) for different bars  in photoelectrons (p.e.) per a minimum ionizing particle (MIP) is plotted in Fig.~\ref{fig:bar4}. 
The readout was performed using MRS APDs and the light yield values have been corrected for dark pulses, optical crosstalk and afterpulsing in the photodiodes. 
The 1~cm wide bar produces the highest light yield when compared to the wider bars.  The light yield of 15~p.e./MIP in a central point along the 1~cm wide bar at a distance of 8 m from both photosensors drops to  10~p.e./MIP for the 4~cm wide bar. The MIP energy deposit in a bar  is  1.4~MeV and corresponds to a minimum light yield of about 7~p.e./MeV in a 0.7~cm thick bar of 4~cm width. Readout using MRS APDs and MPPCs produced similar results  within the measurement errors as shown in Fig.~\ref{fig:bar2}. 

Time resolution in the central part of the fiber was measured to be $\sigma_t\sim$2.0--2.2~ns for the bars with widths ranging from 1 to 4~cm. We made no time walk correction because the discriminator is fired at the arrival of a first photoelectron pulse. Timing is essentially dominated by the Y11 fiber emission decay time. The decay time of Y11 was measured at the center of a 3.5 m long fiber by fitting the time spectrum of a single photoelectron detected at one fiber end while imposing the condition that a few p.e.s where detected at the opposite end.  The decay time of Y11 fiber was found to be 12$\pm$0.5~ns. Using the light propagation velocity in the fiber the spatial resolution along the bar is estimated to be  $\sigma_x\sim$32--35~cm. 

\section{Conclusion}
Scintillator bar readout using a long WLS fiber and multi-pixel photodiodes was studied for a possible usage in large size neutrino detectors. The relative light transmittance for the fiber was measured and compared to the spectral sensitivity of the multi-pixel photodiodes. An attenuation length of 4.2~m at $\sim$520~nm enables us to obtain a good  signal at a distance of 8~m from an  ionization point. 

Extruded scintillator bars  were read out using a 16~m long Y11 fiber and multi-pixel photodiodes (MRS APDs and MPPCs). The minimum  light yield from both fiber ends was measured to be $\sim$10~p.e./MIP for the 4~cm wide bar and increased to 15~p.e./MIP for the 1~cm wide bar of the same thickness of 0.7~cm. 

The tests have demonstrated that reading out both ends of 16~m long scintillator bars with a single Y11 WLS fiber and multi-pixel photodiodes can provide a high detection efficiency for minimum ionizing particles.

\section{Acknowledgements}
This work was supported in part  by  the ``Neutrino Physics'' Program 
of the Russian Academy of  Sciences, by the RFBR (Russia)/JSPS (Japan) 
grant 11-02-92106 and by Science School grant 65038.2010.2.
 


\begin{thebibliography}{}
\bibitem{minos} D. G. Michael et al, Nucl. Instr.  Meth. A596 (2008) 190. 
\bibitem{scibar}K. Nitta et al., Nucl. Instr.  Meth. A535 (2004) 147.
\bibitem{minerva}D.~Drakoulakos et al., hep-ex/0405002; K.S.~McFarland, Nucl. Phys. Proc. Suppl. 159 (2006) 10. 
\bibitem{nova}D. Ayres   et al., hep-ex/0210005; hep-ex/0503053.
\bibitem{iss_det}T.~Abe et al., JINST, 4 (2009) T05001; arXiv:0712.4129. 
\bibitem{renker}D.~Renker and E.~Lorenz, JINST 4 (2009) P04004.
\bibitem{nd280}Yu. Kudenko,  Nucl. Instr.  Meth. A598 (2009) 289.
\bibitem{t2k}T2K collaboration, K.~Abe et al., 
arXiv:1106.1238v2 [physics.ins-det].
\bibitem{mppc} MPPC  (Multi-Pixel Photon Counter) is the trademark of HAMAMATSU PHOTONICS K. K. for  multi-pixel Geiger mode  avalanche photodiodes; 
M.~Yokoyama  et al.,  Nucl. Instr.  Meth. A610 (2009) 128; 
A.~Vacheret et al., arXiv:1101.1996 [physics.ins-det].
\bibitem{mrs} MRS APD is a trademark of the Center of Perspective
Technologies and Apparatus (CPTA, Moscow) for  multi-pixel   photodiodes  with a  Metal--Resistor--Semiconductor layer  structure; http://www.cpta-apd.ru 
\bibitem{musienko}Y.~Musienko et al., Nucl. Instr.  Meth. A567 (2006) 57.
\bibitem{kuraray} Kuraray Co., Methacrylic Resin Division,  Tokyo, Japan.
\bibitem{extrusion}Y.~Kudenko et al., Nucl. Instr.  Meth. A469 (2001) 340.
\bibitem{smrd}O.~Mineev et al., Nucl. Instr.  Meth. A577 (2007) 540. 

\end{thebibliography}
\end{document}